\documentclass[aps,prx,preprint,groupedaddress]{revtex4-2}

\usepackage{graphicx,float}
\usepackage[normalem]{ulem}
\usepackage{xcolor}
\usepackage[euler]{textgreek}
\usepackage[version=4]{mhchem}
\usepackage{braket}
\usepackage{hyperref}
\hypersetup{
    colorlinks=true,
    linkcolor=blue,
    urlcolor=blue,
    citecolor=blue,
}
\usepackage[capitalise]{cleveref}
\usepackage{siunitx}
\usepackage{multirow}
\usepackage{textcomp}

\begin{document}
\title{Microscopic study of optically-stable, coherent color centers in diamond generated by high-temperature annealing}

\author{King Cho Wong$^{1\S}$}
\author{San Lam Ng$^{1\S}$}
\author{Kin On Ho$^{1\S}$}
\author{Yang Shen$^{1,2}$}
\author{Jiahao Wu$^{1,2}$}
\author{Kwing To Lai$^1$}
\author{Man Yin Leung$^{1,2}$}
\author{Wai Kuen Leung$^{1,2}$}
\author{Durga Bhaktavatsala Rao Dasari$^{3}$}
\author{Andrej Denisenko$^{3}$}
\email{a.denisenko@pi3.uni-stuttgart.de}
\author{J\"{o}rg Wrachtrup$^{3}$}
\author{Sen Yang$^{1,2}$}
\email{phsyang@ust.hk}

\affiliation{$^1$ Department of Physics, The Chinese University of Hong Kong, Shatin, New Territories, Hong Kong, China}
\affiliation{$^2$ Department of Physics and the IAS Centre for Quantum Technologies, The Hong Kong University of Science and Technology, Clear Water Bay, Kowloon, Hong Kong, China}
\affiliation{$^3$ 3rd Institute of Physics, University of Stuttgart and Institute for Quantum Science and Technology (IQST), Pfaffenwaldring 57, D-70569, Stuttgart, Germany}
\affiliation{$^{\S}$ These authors contributed equally to this work}

\begin{abstract}
Single color centers in solid have emerged as promising physical platforms for quantum information science. Creating these centers with excellent quantum properties is a key foundation for further technological developments. In particular, the microscopic understanding of the spin bath environments is the key to engineer color centers for quantum control. In this work, we propose and demonstrate a distinct high-temperature annealing (HTA) approach for creating high-quality nitrogen vacancy (NV) centers in implantation-free diamonds. Simultaneously using the created NV centers as probes for their local environment we verify that no damage was microscopically induced by the HTA. Nearly all single NV centers created in ultra-low-nitrogen-concentration membranes possess stable and Fourier-transform-limited optical spectra. Furthermore, HTA strongly reduces noise sources naturally grown in ensemble samples, and leads to more than three-fold improvements of decoherence time and sensitivity. We also verify that the vacancy activation and defect reformation, especially H3 and P1 centers, can explain the reconfiguration between spin baths and color centers. This novel approach will become a powerful tool in vacancy-based quantum technology.

\end{abstract}

\maketitle

\section{Introduction}
Colour centers such as point defects and dopants in solid hosts have shown advantages such as long decoherence time, excellent optical properties, good addressability and high-fidelity controls \cite{Wrachtrup06, Atature2018Review, Awschalom2018Review}. These features have led to the realization of a wide range of quantum applications. The high-quality solid host materials and mature nano-fabrication techniques have been well developed, thus opening the room for the fabrication of sophisticated devices. For example, to have the best quantum properties, substrates with high purity are preferred, such as 3-ppb-nitrogen diamonds. In those samples, however, the existing color center density is extremely low as the CVD as-grown NV yield is below one percent. Although color centers can be created spontaneously during host growth, they are difficult to create post-growth. Up to now, ion-implantation-based methods have been widely used. These methods, however, introduce unwanted damages along the beam paths. Except for a few special systems, which benefit from their own unique lattice structure (e.g., silicon vacancy in diamond, which has a mirror-symmetric structure \cite{Tamura2014SiV}), unwanted damages can significantly degrade the properties of solid-state qubits, even after various post-fabrication treatments like thermal annealing. This problem hinders further developments in the field. In this work, we implement a high-temperature annealing (HTA) approach to create color centers, by using both vacancy activation and defect-species conversion processes that take place at high temperatures, without introducing unwanted defects or deteriorating the quality of qubits. The HTA method significantly improves the decoherence time by reduces existing noise source in the host. We also identify the defect-species conversion processes using coherence time measurement and double electron-electron resonance (DEER). This implantation-free method will accelerate progress in quantum technology based on solid-state qubits.

We demonstrated the HTA method on nitrogen vacancy (NV) centers in diamond, a model system that has suffered seriously from the aforementioned problems. The NV center is also a sensitive sensor that we use to microscopically characterise the possible damage this technique introduces. The NV center, which consists of a nitrogen dopant and a nearby vacancy, is a negatively charged deep-level defect in diamond \cite{Wrachtrup06, Hollenberg13}. It has an effective spin of 1 and features such as high-fidelity optical spin initialization, control and readout, as well as long coherence time and excellent optical properties at low temperatures \cite{DohertyNJP2011, Wrachtrup06, Atature2018Review, Awschalom2018Review}. Notably, optically detected magnetic resonance (ODMR) can be observed even at ambient conditions. It has been used in all three major directions in quantum technologies: quantum computing, communication and sensing. Its unique optical properties such as spin selective transitions at low temperatures provide opportunities for quantum coupling with photons and quantum entanglement over a long distance \cite{Lukin2010Entanglement, Hanson2013Entanglement, Yang2016NPhotonics}. Scaled-up quantum nodes have been realised with the NV center due to the addressability of nearby nuclear spins \cite{TimTaminiau2019TenQubits, TimTaminiau2019Imaging}. Furthermore, the NV center is sensitive to environmental parameters such as magnetic and electric fields, temperature and stress \cite{Wrachtrup06, Hollenberg13, SY2021JAP, SY2019Science}. While this makes the NV center a powerful sensor, it puts strict requirements for the surroundings. In fact, any noise or instability (and especially nearby paramagnetic defects) can strongly degrade its excellent as-fabricated properties.

Nitrogen is one of the most common dopants in diamonds, but only a tiny fraction of substitutional nitrogen atoms exist in the form of NV centers. Most nitrogen atoms occur in other forms, including P1 (\ce{^{14}N} or \ce{^{15}N}) centers, N3 (3N+V) and H2 (NVN$^{-}$) defects. These paramagnetic defects can induce significant noise in the NV center. Besides being created naturally during diamond growth, NV centers can be created via several post-growth approaches. In general, the presence of both a nitrogen dopant and a vacancy is necessary to create an NV center. Nitrogen dopants (i.e. P1 centers) can be incorporated during diamond growth or by implanting a diamond substrate with nitrogen ions or molecules, while vacancies (i.e. GR1 defects) can be induced by implanting ions (e.g. nitrogen, carbon or helium) or by laser or electron-beam (e-beam) irradiation. When the substrate is annealed above 800\textsuperscript{o}C, vacancies in the lattice becomes mobile and can be captured by a nitrogen dopant to form an NV center \cite{JMeijer2005Implantation, JMeijer2021Review, Collins1980Vacancy, Luhmann2018Screening, Alekseev2000Transformation, Alison1994Nitrogen}. Many methods of NV center creation have been developed by combining different techniques, including ion implantation, delta doping and laser or e-beam irradiation \cite{JMeijer2005Implantation, Chen2017LaserWriting, Hanson2019MembraneEbeam, LaserWriting}. However, these methods introduce unavoidable damages in the diamond lattice: laser irradiation or ion implantation can cause voids or broken bonds along the beam path, and delta doping or nitrogen implantation can induce P1 centers (as the density of NV centers is usually orders of magnitude smaller than that of P1 centers). These damages, in turn, may induce magnetic and electric noise, as well as charge state instability in nearby NV centers. This further reduces the decoherence time, degrades the optical fine structure (even at low temperatures) and leads to charge state switching from the preferred NV$^-$ to NV$^0$ state \cite{Manson2005PhotoIonization}. Post-processing methods including annealing the sample at 800$\sim$1200\textsuperscript{o}C and improving surface terminations have been introduced to reduce these effects \cite{Naydenov2010Increasing, Fu2010SurfaceTermination, Lukin2014Implantation}. Up till now, laser irradiation method can create NV centers with linewidth from tens MHz to a few hundreds MHz \cite{Chen2017LaserWriting}, while 1200\textsuperscript{o}C annealing after the ion implantation can reach the linewidth within tens MHz to GHz range \cite{Lukin2014Implantation}. However, it is still challenging to create NV centers with most of them having properties on a par with those of CVD as-grown NV centers. Furthermore, the above methods usually require special advanced facilities and sophisticated processing, which can damage fragile samples like diamond membranes or devices.

This implantation-free color center creation method is based on the complicated physical and chemical processes of defects and vacancies under high temperature in a diamond. The HTA method has its working temperature around 1700\textsuperscript{o}C. This high temperature significantly changes the physics picture during annealing, as two new processes emerge and dominate all over the host: the activation and migration of vacancies, and the reformation of defects. High temperature activates significant and complex vacancy dynamics inside the host material, as well as large aggregations of nitrogen dopants even in low-nitrogen-concentration samples. Vacancies activated at sample surfaces can also migrate at high temperatures to the bulk \cite{Collins1980Vacancy}. Moreover, at 1500\textsuperscript{o}C, nitrogen can diffuse, albeit slowly. Therefore, at 1700\textsuperscript{o}C, both nitrogen and vacancies will be mobile in the diamond lattice, forming various defects, such as NV centers, as previous studies have shown that a mobile vacancy can be captured by a nitrogen \cite{JMeijer2005Implantation, JMeijer2021Review, Collins1980Vacancy, Luhmann2018Screening, Alekseev2000Transformation, Alison1994Nitrogen, Goss2004Interstitial}. Here we have obtained high-quality NV centers in diamonds with a wide range of samples with varying nitrogen concentrations. Using the generated NV centers as probes of the local environment, we have verified that this novel technique does not introduce extra noise sources microscopically. Furthermore, complicated defect reformation processes taking place under such high temperature greatly reduce paramagnetic noise sources. Therefore, ensemble NV centers created in high-nitrogen-concentration samples have their spin performance significantly improved.


\section{The Creation of NV centers}
\begin{figure}[h!]
\includegraphics[width=12cm]{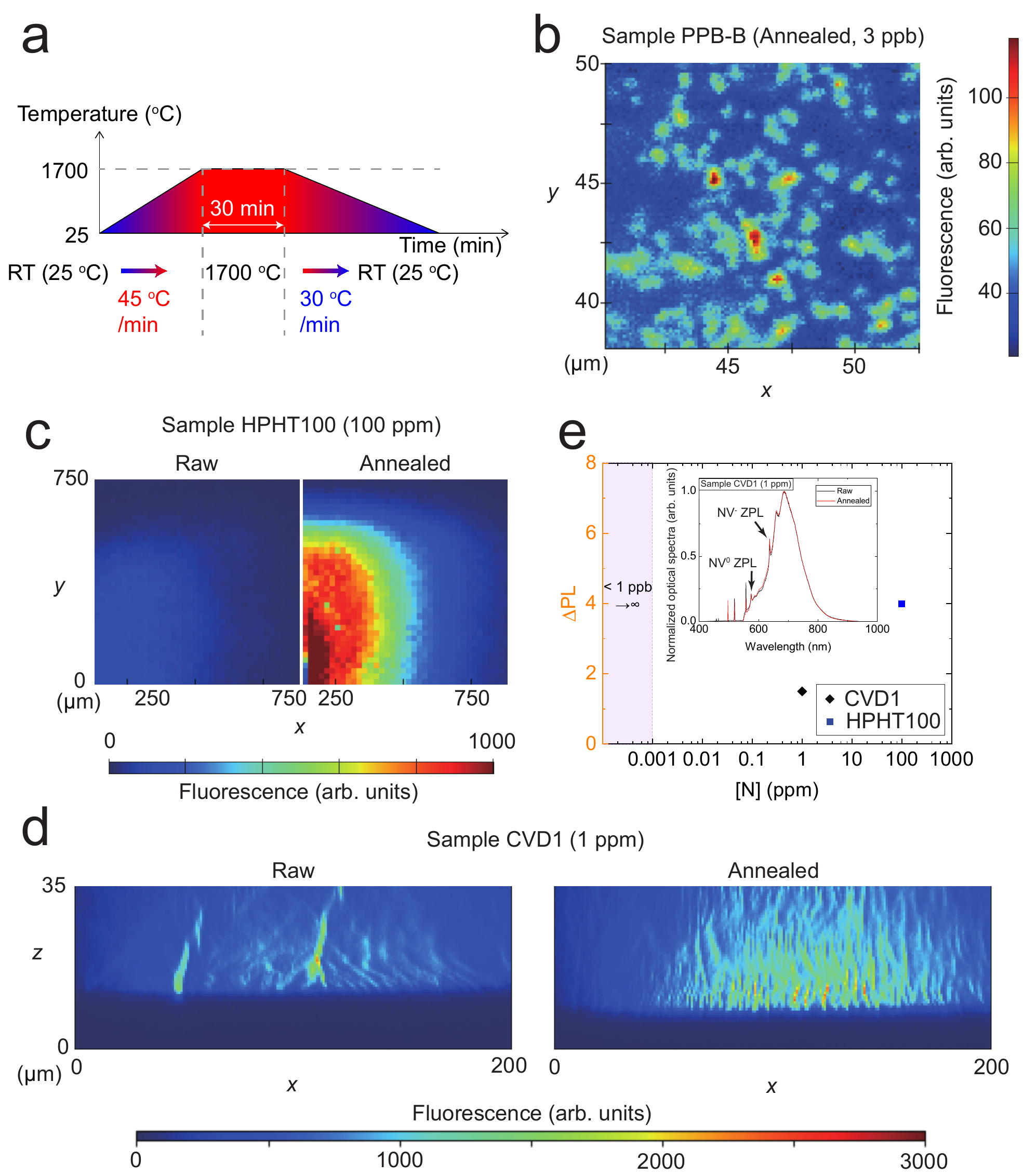}
\centering
\caption{\textbf{The protocol of the HTA method and the NV center creations in different samples.} \textbf{a}, The time sequence of HTA. The peak temperature of 1700\textsuperscript{o}C is held for 30 mins in a high-temperature furnace under a vacuum. \textbf{b}, A confocal image ($y-x$ scan) of the ultra-low [N] sample PPB-B after annealing, revealing noticeable single NV center creations. \textbf{c}, A confocal image ($y-x$ scan) of raw and annealed sample HPHT100 with fixed measurement parameters. Ensemble NV centers are created in this sample. \textbf{d}, A confocal image ($z-x$ scan) of raw and annealed sample CVD1 with fixed measurement parameters. The color scale denotes the PL counts, which provide a fair comparison between the same sample before and after annealing. \textbf{e}, The increase in PL of ensemble samples CVD1 and HPHT100 after the annealing treatment. Normalized optical spectra of sample CVD1 under 520-nm excitation are shown in the inset.}
\label{fig1}
\end{figure}

The sequence of the HTA is shown in \cref{fig1}a. The temperature is ramped up to 1700\textsuperscript{o}C at a rate of 45\textsuperscript{o}C/min and held constant for 30 min. Then, the system is then quickly cooled down to the ambient temperature at a rate of -30\textsuperscript{o}C/min. After this, diamond samples are subjected to a boiling acid treatment to remove graphitized layers from the diamond surfaces. This work examines both Type Ib and Type II diamond samples, with nitrogen concentrations that are typical for applications in quantum technologies. A total of four different categories of diamonds have been investigated, including two samples with nitrogen concentration [N] below 3 ppb (sample PPB-B is a bulk diamond, and sample PPB-M is a 20 {\textmu}m thick membrane), sample CVD1 ([N] $=1$ ppm) and sample HPHT100 ([N] $=100$ ppm). ([ ] denotes concentration.)


We have observed the formation of single NV centers in samples PPB-B and PPB-M, and a noticeable increase in the photoluminesce (PL) of NV centers in two ensemble samples, CVD1 and HPHT100, after HTA. For ultra-pure samples PPB-B and PPB-M with [N] of 3 ppb as grown, we could not detect any single NV center in confocal microscopy scans, since the NV center created during the growth is tiny in both percentage (below one percent) and in density (usually below 1 per 100 \textmu m$\times$100 \textmu m area). This is a serious problem for quantum information applications, as there is nearly no as grown NV centers to start with and the optical properties of implanted NV centers can be compromised. But after HTA, NV centers are formed with a concentration $> 1$ center per 5 \textmu m$^{3}$, (see the confocal image in \cref{fig1}b), an ideal concentration required for applications in quantum computing and communication. This density corresponds to NV center concentration of 0.5ppb. As the nitrogen concentration is below 3ppb, the NV center yield is over 17\% yield, which is orders of magnitude than the CVD as-grown yield.

For samples with higher nitrogen concentrations, shown in \cref{fig1}c,d, we compare the confocal images of the raw and annealed ensemble samples (samples HPHT100 and CVD1) under the same experimental conditions. We clearly observe an increase in the PL after annealing. In particular, the depth scan ($z-x$ scan), as shown in \cref{fig1}d, clearly shows that the change is uniform over the sample, and cannot be related to artefacts, for example due to surface contamination. These results show that HTA increases NV center concentration [NV] in the bulk. The increased PL of samples shown in \cref{fig1}e, clearly reveals a significant increase of [NV]. Sample HPHT100 shows a $\sim$4 times increase while sample CVD1 shows a $\sim 1.5$ times increase. Furthermore, we see that a preferential NV center orientation in sample CVD1 found before is completely lost after the thermal annealing \cite{Osterkamp2019Engineering}. This loss of preferential NV center orientation in the ensemble can be attributed to the symmetry of the four possible lattice axes (see Supplementary Note III). Also, there might be another re-orientation process that NV centers can undergo at 1050\textsuperscript{o}C as shown in \cite{Chakravarthi2020Window}. In the inset of \cref{fig1}e we show that under 520-nm excitation, the ratio between NV$^{0}$ to NV$^{-}$ states remains unchanged for sample CVD1. This could be attributed to the fact that the Fermi level remains unchanged even after the thermal treatment. It is known that divacancy (V$_{2}$) defects, which can be annealed out at 800\textsuperscript{o}C, are deep acceptors than NV centers, and hence their reduction should convert NV$^{0}$ to NV$^{-}$ \cite{Deak2014Formation}. But for sample CVD1, the increase in PL is from the increase of [NV], and not due to the conversion from NV$^{0}$ to NV$^{-}$. In addition, within our sensitivity, no optical signal from GR1 defects \cite{Zaitsev2001Optical} is observed in any case, concluding that the residual vacancies from crystal grown are low, and the vacancies involved to create NV centers are mostly thermally activated. Due to the increase in [NV], we also improve the sensitivity of the NV ensemble sensor (details below).

\begin{figure}[h!]
\includegraphics[width=12cm]{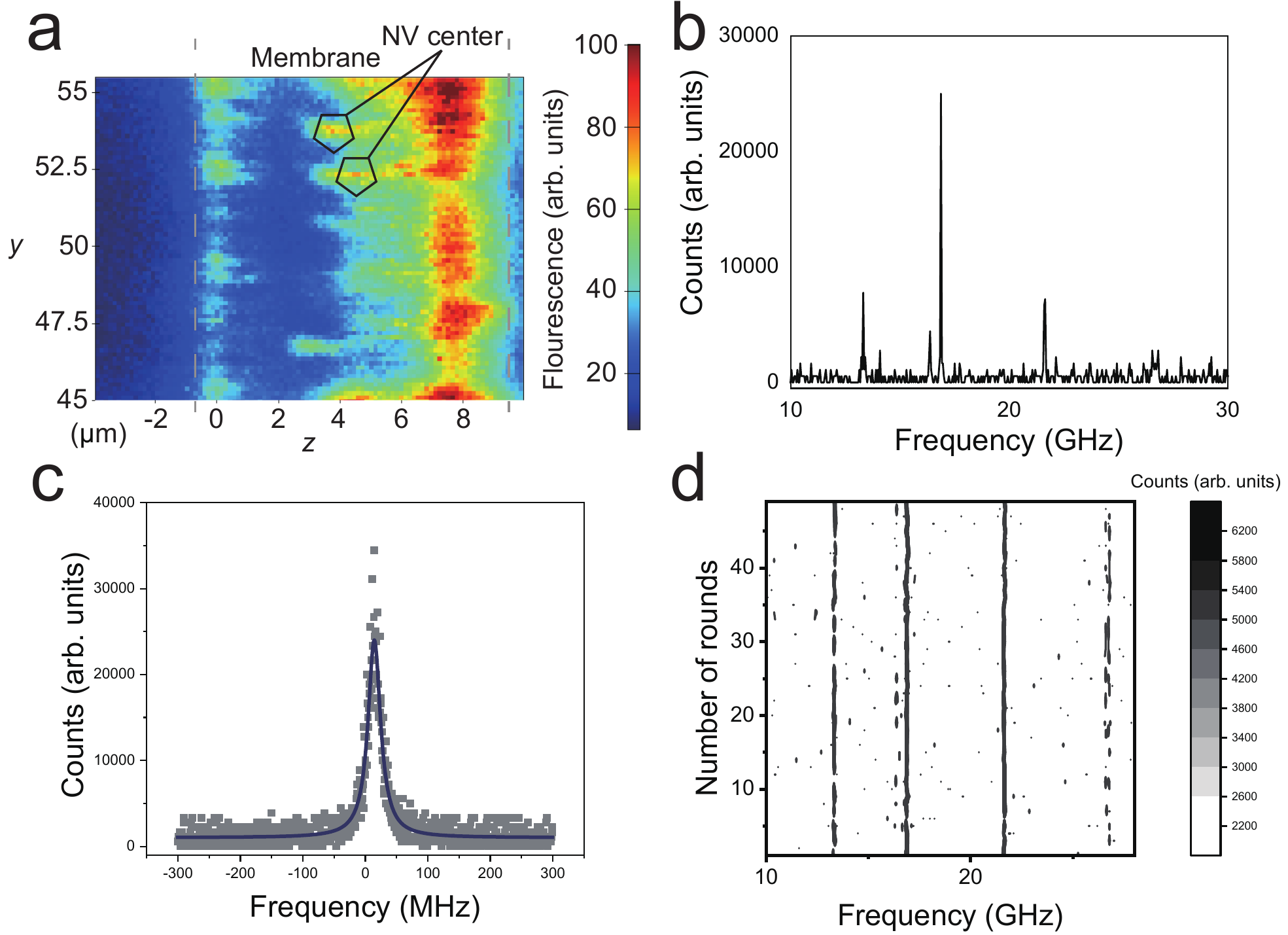}
\centering
\caption{\textbf{Optical properties of created NV centers in a 20 {\textmu}m thick diamond membrane.} \textbf{a}, Confocal image of the cross-section of the membrane. The $z$-axis marks the appearance depth (the correction fact due to the effective numerical aperture of the objective and refractive index of diamond is around 2.7). \textbf{b}, A PLE spectrum showing the 6 sub-levels of optical excited states. \textbf{c}, A zoom-in of spectrum line which has the Fourier transform-limited linewidth of 24 MHz. \textbf{d}, 40 rounds of repetitive PLE scan. This confirms the spectral stability of the sample.}
\label{fig2}
\end{figure}

\section{Improved optical and spin properties}

We now proceed to investigate the optical and spin properties of newly created NV centers. One of the unique features of the NV center is its optical excited-state structure with 6 sub-levels split by the spin-orbital coupling at low temperatures. These spin-selective transitions allow spin-photon entanglement, high-fidelity single shot readouts and coherent storage of photon states, which allow the development of promising devices for quantum communications and computing \cite{Lukin2010Entanglement, Hanson2011Readout, Hanson2013Entanglement, Yang2016NPhotonics}. These optical transitions are sensitive to local electric field noise with the electric field susceptibility $\sim1$ MHz/(V/cm) \cite{Yao2021Optically}, making them vulnerable to the surrounding electric noise, for example due to the charge defects in the close vicinity of NV centers. Moreover, laser excitation can cause further charge redistribution among those defects and deteriorate the spectral stability of NV centers. Experimentally, these effects have been observed in the form of spectral diffusion, spectral jumps, a broadening of spectral lines, the disappearance of spin-selective transitions, and fast charge state ionization. The unwanted defects created during the traditional ion implantation methods for the creation of NV centers usually deteriorate the resonant optical properties of the NV centers at low temperatures. With the HTA approach here, in both the bulk sample (sample PPB-B) and a 20 {\textmu}m diamond membrane (sample PPB-M), all created NV centers possess the same spectral stability similar to the damage-free natural samples. In \cref{fig2}, we show NV centers created in the membrane sample PPB-M. As shown in the confocal image in \cref{fig2}a, [NV] is at least 1 center per 5 {\textmu}m$^3$. At 7 K, the six optical excited-state sub-levels can be clearly resolved in the photoluminescence excitation (PLE) spectrum shown in \cref{fig2}b. The linewidth is $24$ MHz (\cref{fig2}c) which is Fourier transform limited. We perform over 40 rounds of repetitive PLE scans (\cref{fig2}d) and the resulting spectra are found to be stable, with no spectrum diffusion or charge state switching. All $18$ NV centers examined in this sample have shown similar PLE spectra and stability with linewidth below 50MHz (see Supplementary Note III for additional data). Moreover, these sensitive PLE measurements prove that the HTA does not create, microscopically, any additional electrical noise around NV centers. To build a high performance diamond quantum node, NV center has to be spectral narrow, optical stable, and with low strain. Photonics structures such as solid immersion lens require the depth of NV center to be within a few to ten microns. It is quite hard as there is nearly no as grown NV centers in [N]$\sim$ppb diamond bulks. This HTA method is one rare method that can create NV centers with stable and uniform optical properties in large quantities inside otherwise nearly empty diamond host.

\begin{figure}[h!]
\includegraphics[width=12cm]{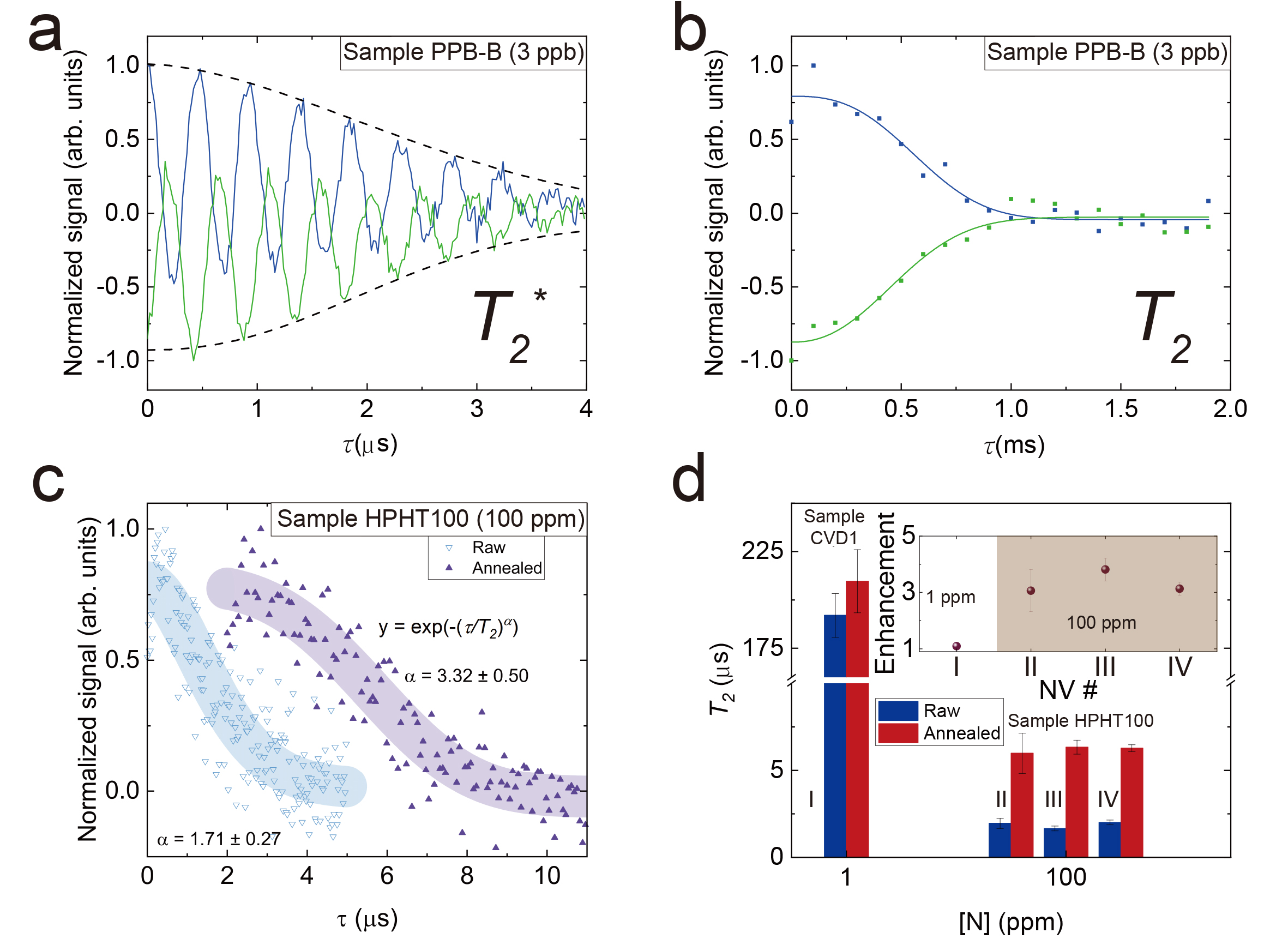}
\centering
\caption{\textbf{The universal enhancement of electron spin properties in different types of samples.} \textbf{a, b}, The FID and Hahn echo signals from a single NV center created by this method in diamond sample PPB-B. \textbf{c}, The $T_{2}$ measurements of sample HPHT100 before and after annealing in the same sample spot (Spot IV in (d)) . \textbf{d}, The summary of $T_2$ time of the ensemble samples CVD1 and HPHT100, providing a direct comparison of the raw and annealed data. Inset summarises the enhancement of the $T_{2}$ time. The label I corresponds to a randomly chosen spot in sample CVD1. The labels (II, III and IV) correspond to three randomly chosen spots in sample HPHT100. For sample PPB-B, the external magnetic field strength was $\SI{546}{G}$. For sample CVD1, the external magnetic field strengths were $\SI{80}{G}$ before the annealing and $\SI{79}{G}$ after annealing. For sample HPHT100, the external magnetic field strengths were $\SI{49}{G}$.}
\label{fig3}
\end{figure}

We now characterise, microscopically, the electron spin properties of the created NV centers both through the free induction decay (FID) and Hahn echo methods at ambient conditions. The two spin-baths (spin noise sources) that generally influence the NV center spin-coherence times are $^{13}$C nuclear spin bath and the P1 electron spin bath. We choose three samples with different spin-noise environments: (i) where the $^{13}$C nuclear spin bath dominates, which is the case for sample PPB-B, with [N] = 3 ppb and natural-abundance $^{13}$C isotope, (ii) where P1 electron spin-bath dominates, which is the case for sample HPHT100 with [N] = 100 ppm and (iii) where both noise sources have similar contribution, which is the case for sample CVD1 with [N] = 1~ppm and natural-abundance $^{13}$C isotope. In \cref{fig3}, we plot the decay of the NV center spin coherence for these three cases for FID and Hahn echo sequences. To extract the $T_2$ time, we use a stretched exponential function $e^{- \left( \frac{t}{T_2} \right)^\alpha}$ to fit the Hahn echo data \cite{HahnExponent}. As shown in \cref{fig3}a,b, for NV centers created in samples PPB-B (case (i)), $T_{2}^{*} \approx 3.04 \pm 0.91$ \textmu s, and $T_{2} \approx 0.66 \pm 0.46$ ms. These values are similar to the naturally created NV centers whose spin lifetimes are solely limited by the $^{13}$C spin bath which does not change even after the HTA. This result also proves that for low [N] samples the HTA method does not microscopically introduce any extra paramagnetic noise sources around single NV centers.

For samples CVD1 and HPHT100, the spin-decay behaviour for Hahn echo sequence from which we obtain $T_2$ time is shown in \cref{fig3}c,d.  We discuss the FID behaviour for obtaining $T_{2}^{*}$ in Supplementary Note III. Both samples show an improvement in $T_2$ time after annealing. Sample HPHT100 has shown an increase by a factor $3.3$ of the spin coherence lifetime, while for sample CVD1 with much less [N] and severely influenced by its $^{13}$C nuclear spin bath, the increase in $T_2$ time is nominal i.e., only by a factor $1.1$. Furthermore as shown in \cref{values}, for both samples HPHT100 and CVD1, the exponent $\alpha$ in the stretched exponential fitting $e^{- \left( \frac{t}{T_2} \right)^\alpha}$ increases after annealing. This is particularly pronounced for sample HPHT100 which also showed an enhanced spin coherence time $T_2$. As shown previously in a theoretical work \cite{HahnExponent}, this exponent could reach a value $3$ when the correlation time of spin-noise is approaching infinite. In the other extreme limit it will take the value $1$ when the correlation time is approaching $0$. Therefore, the enhancement of the spin-coherence time and the increase in the exponent together indicate that the HTA greatly influences the noise spectrum of the spin bath. We would like to note that this HTA is one among the few methods that can reliably create spin-qubits and at the same time an enhanced spin coherence time. The increased [NV] and an improved decoherence time will double the effect on the improvement of the sensitivity estimates \cite{Barry2020Sensitivity, Degen2017Quantum}, given by

\begin{equation}
\eta_{ensemble} \approx \frac{\hbar}{\Delta m_{s} g \mu_{b}}\frac{1}{\sqrt{[NV] \tau}}.
\end{equation}
Here $\Delta m$ is the change of spin quantum number, $g \approx 2$ for the NV center and $\mu_{b}$ is the Bohr magneton. Considering sample HPHT100, the increase of [NV] and the $T_{2}$ time $\tau$ give about 3.6 times improvement on sensitivity over naturally grown samples.

\begin{table}[h!]
\centering
\begin{tabular}{||c|c|c|c|c|c|c|c|c||}
\hline
Label & [N] & [NV] & $\alpha_r$ & $\alpha_a$ & [P1]$_{r}^{1}$ & [P1]$_{a}^{1}$ & [P1]$_{r}^{2}$ & [P1]$_{a}^{2}$
\\ \hline
Sample CVD1 Spot I & 1 & $\sim0.01$ & $1.48 \pm 0.18$ & $1.60 \pm 0.28$ & 0.20$\pm$0.04 & 0.16$\pm$0.04 & 0.61$\pm$0.15 & 0.54$\pm$0.15
\\ \hline
\multirow{3}{*}{Sample HPHT100 Spot II-IV} & \multirow{3}{*}{100} & \multirow{3}{*}{$\sim0.001$} & $1.40 \pm 0.29$ & $1.82 \pm 0.71$ & 75.6$\pm$20.5 & 6.5$\pm$4.0 & 81.7$\pm$20.3 & 26.5$\pm$7.4
\\ \cline{4-9}
& & & $1.27 \pm 0.20$ & $2.66 \pm 0.54$ & 52.6$\pm$7.8 & 8.0$\pm$3.6 & 96.2$\pm$20.7 & 25.1$\pm$5.2
\\ \cline{4-9}
& & & $1.71 \pm 0.27$ & $3.32 \pm 0.50$ & 67.1$\pm$9.5 & 19.5$\pm$4.2 & 79.7$\pm$16.7 & 25.3$\pm$5.1
\\ \hline
\end{tabular}
{\raggedright
 \\All data in [ ] are in ppm unit.\\
 \textsuperscript{1} Determined from DEER spectra.\\
 \textsuperscript{2} (For reference) Determined from $T_{2}$ time \cite{P1ConcentrationfromT2}, under the assumption that the leading decoherence source is the P1 spin bath. This assumption is valid for sample HPHT100 with 100 ppm nitrogen, but may become improper for sample CVD1 which is significantly influenced by the $^{13}$C nuclear spin bath.\\
 \textsubscript{r} Raw.\\
 \textsubscript{a} Annealed. \par}
\caption{\textbf{The exponent $\alpha$ derived from Hahn echo measurements and [P1] derived from DEER spectra for the two samples before and after annealing.} One spot is tracked in sample CVD1 while three spots are tracked in sample HPHT100 for comparison (the same spots as in Fig.3). The exponent $\alpha$ increases slightly for sample CVD1 whose major noise sources are both $^{13}$C spin bath and P1 centers, while clear changes are observed in sample HPHT100 whose major noise source is the P1 spin bath, suggesting significant changes in the spin bath environment for sample HPHT100. [P1] is low in the dilute sample and the change after HTA is tiny, while the change is significant in a denser sample. In both cases, only a small fraction of substitutional nitrogens are in the form of NV centers. The estimated [NV] are also shown for the two samples.}
\label{values}
\end{table}

\section{the microscopic mechanism of color center creation and coherence improvement}

To understand the observed modification of the noise environment (spin bath reconfiguration), we would like to analyse the role of P1 centers. As the concentration of [P1] is at least an order of magnitude higher than [NV] (shown in \cref{values}), just converting a small proportion of P1 centers into NV centers cannot explain the significant increase in the observed spin coherence times. More complicated defect reformation processes are needed to be taken into account. It was shown that \ce{NVN^{0}}/\ce{NVN^{-}} (i.e. H3/H2 defects) can be formed under thermal annealing \cite{Collins1980Vacancy, Collins2005High, Collins1978Migration, Luhmann2018Screening, Alison1994Nitrogen, Lawson1992The,  Alekseev2000Transformation}. Therefore, during the annealing process, some of the P1 centers might combine with vacancies or even NV centers, forming \ce{NVN^{0}}/\ce{NVN^{-}}. Such combinations would favour an extension of the decoherence time of the remaining NV centers, as two P1 centers would be eliminated to form a \ce{NVN^{0}}/\ce{NVN^{-}}. Notably, \ce{NVN^{0}} has no electron spin, while \ce{NVN^{-}} posses a 1/2-spin. Effectively, two 1/2-spins (i.e. two noise sources) would be eliminated and replaced by either a zero spin or one 1/2-spin. Due to this we expect that the overall spin noise is significantly reduced.


\begin{figure}[t]
\includegraphics[width=10.5cm]{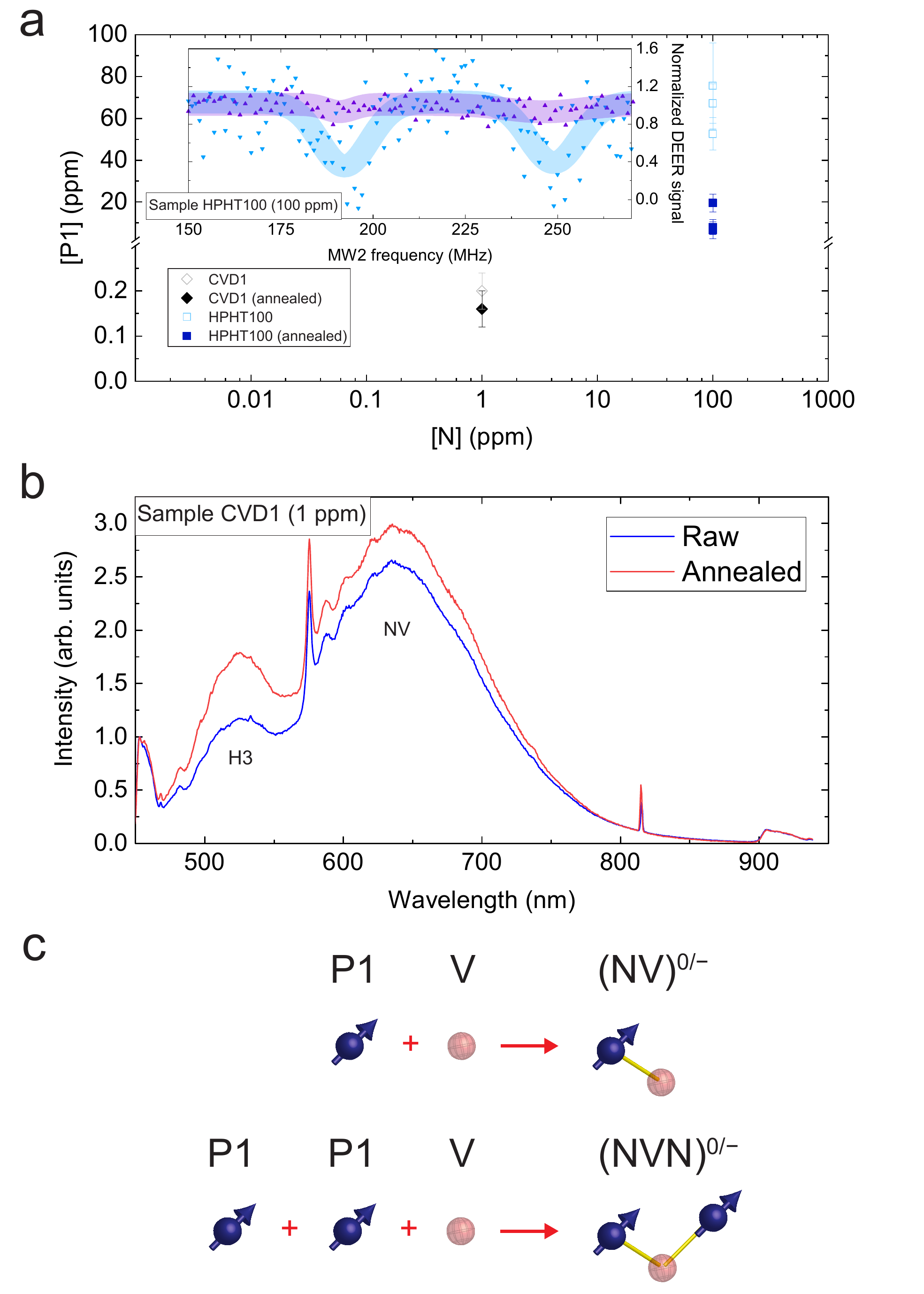}
\centering
\caption{\textbf{The change in the spin bath after HTA and the related mechanism.} \textbf{a}, P1 concentration of all samples. Inset shows a set of DEER spectra of sample HPHT100 before and after annealing. An explicit decrease in contrast is observed. \textbf{b}, Optical emission spectra of sample CVD1 under 405-nm excitation. \textbf{c}, Schematics of the combination of P1 centers and vacancy into NV center and \ce{NVN^{0}}/\ce{NVN^{-}}.}
\label{fig4}
\end{figure}

To verify this hypothesis, the concentration of various dopants and defects has to be measured. Electron paramagnetic resonance (EPR) and absorption spectrum have macroscopically shown evidence on defects reconfiguration in the whole sample. However, samples with high dopant concentrations usually do not have a uniform dopant density, because experimental conditions drift during the long period of sample growth. Therefore, the averaged density of a spin bath over the whole sample may not relate to the spin bath in a given diffraction-limited size spot. Here, one spot is tracked in sample CVD1 while three spots are tracked in sample HPHT100 for comparison. We use near-field methods like DEER to probe the local concentrations of paramagnetic defects and PL spectra to estimate concentrations of spinless defects.

P1 centers have unique ESR spectra, due to their hyperfine coupling with nitrogen, and can be clearly isolated from other peaks in the spectra \cite{Viktor2016Determination}. By adopting the protocol of Ref.\cite{Li2021Determination}, we derive the P1 concentration, as shown in \cref{fig4}a. An observable drop in [P1] can be seen in both samples CVD1 and HPHT100. The inset presents selected DEER spectra of sample HPHT100. An explicit decrease in contrast is observed. A summary of the defects concentration is presented in \cref{values}. Note that [P1] shows an obvious drop after annealing, [P1] of the annealed sample CVD1 is reduced to $\sim 84\%$ of its original value, while [P1] of the annealed sample HPHT100 is reduced to $\sim 17\%$ on average. The change in [P1] seems to correlate with the enhancement of $T_{2}$ time, meaning that the large reduction of [P1] of the sample HPHT100 contributes to the significant increase in $T_{2}$ time. In short, under proper normalization (see Supplementary Note II), our data confirm, microscopically, a decrease in [P1] in our samples after annealing.

We also estimate the order of magnitude of [NV] from the PL intensity, as shown in \cref{values}. The amount of P1 centers which disappeared is much more than the increase in NV center concentration. Therefore, it hints that a large amount of P1 centers are converted into other forms of defect during the thermal process. One of the candidates is H3 defects which can be formed via P1 and vacancy \cite{Davies1977Charge, Jones1993Theory, Alison1994Nitrogen, Deak2014Formation, Collins1980Vacancy, Pinto2012On, Collins1980Vacancy, Collins2005High, Collins1978Migration, Luhmann2018Screening}. The H3 is a spinless defect, but its PL emission consists of zero phonon line at 503.2-nm and phonon sideband can be observed in the PL spectra under 405-nm excitation, as presented in \cref{fig4}b and Supplementary \cite{Zaitsev2001Optical}. The integrated H3 PL increases about 2 times after the annealing treatment, which confirms an increase in [H3]. In \cref{fig4}c, we present the schematic drawing of the creation mechanism of NV and H3 centers. Further \textit{ab initio} simulations may reveal a more detailed insight into the involved processes \cite{Galli2005}.

Recently, several works have demonstrated the improvement of diamond particles via rapid thermal annealing (RTA) \cite{Pereira2003The, Gierth2020Enhanced, Torelli2020High, Shames2017Fluence}. By analysing the EPR and absorption spectrum, the RTA method is shown to improve, macroscopically, the quality of diamond particles. Moreover, NV-center driven \ce{^{13}C} hyperpolarization could be enhanced by the RTA in high-NV-concentration 18-\textmu m diamond particles \cite{Gierth2020Enhanced}. We note here that the two major differences between our work and previous works are the observed change in $T_{1}$ time and [P1] of the samples before and after annealing. In our work, there is no pronounced enhancement on the $T_{1}$ time, while in ref.\cite{Gierth2020Enhanced, Torelli2020High, Shames2017Fluence} $T_{1}$ are significantly improved. This may related to the recovery from lattice damages induced by the e-beam irradiation and during nanodiamond fabrication. Moreover, we present clear drops on [P1] same as in ref.\cite{Pereira2003The}, but contradicts to the large raised up in ref.\cite{Gierth2020Enhanced} and almost unchanged (if not decreased) in Ref\cite{Shames2017Fluence, Torelli2020High}. One possible explanation is the difference in protocol. Their RTA protocol is much more rapid in both the holding time and in the ramping up time (a few minutes versus 30 minutes). This affects the dynamics of the defect reconfiguration, resulted in different concentrations. A more intrinsic reason lies on the mechanism of the HTA method. Surface is both the source of vacancy creation and also the sink of vacancy as well as dopant. In contrast to bulk diamonds used in our work, they used 140 nm to 20 \textmu m range diamond particles which have huge surface area/volume ratio as well as small size. These factors strongly change the physical picture, as the diffuse lengths for dopant and vacancies are limited by the particle size and they are strongly influenced by the surface layers.


\section{Outlook}
In summary, we demonstrate the thermal-activation-based implantation-free approach for a robust creation of solid-state spin qubits. We achieve this by simultaneously generating vacancy-related defects through thermal-induced vacancy activation/migration and at the same time removing unwanted paramagnetic defects by defect reformation at high temperature. This HTA technique not only solves the long term challenge for creating qubits without introducing extra noise sources, but also further reduces the original spin-noise environment in the host material. By incorporating nitrogen atoms at targeted positions in a host material via methods like delta doping, one can create high-quality NV centers in those chosen areas by this HTA method. Further theoretical studies on the migration of vacancies and different defect reformation processes may allow one to optimise the parameters of thermal treatment such as its peak temperature and time duration. We envisage that such studies could further benchmark the conditions for a high color center yield with ultra-long spin coherence times and stable optical lines. Therefore, this method is ideal for building high performed optical cavity and nanophotonics devices. This mechanism and line of thought behind this method are general enough, and can be adapted for any vacancy-based solid state system, especially defects in hosts like diamond and SiC which have similar lattice structures. Furthermore, this HTA approach can be extended for industrial applications where a mass production of quantum systems is a prerequisite.

\section{Acknowledgements}

 We would like to thank D. Pan, N. Zhao for fruitful discussions. We thank the technical helps from Y. Chen, Gary C. H. Lai, W. K. Lo, K. Kafenda and Dickon H. L. Ng. K.O.H acknowledges financial support from the Hong Kong PhD Fellowship Scheme. S.Y. acknowledges financial supports from Hong Kong RGC (GRF/14304419). J.W. acknowledges financial supports from ERC grant SMeL, EU Project ASTERIQS, DFG (GRK2642 and FOR2724).

\appendix

\section{Experimental setup}

The samples PPB-B, PPB-M and CVD1 are supplied by Delaware Diamond Knives and Element Six. The sample HPHT100 is supplied by Jinan Diamond Technology Co. The nitrogen concentrations quoted in the work were the upper limit values given by the suppliers.

The thermal treatment of samples was carried out by annealing the diamond samples in a high-temperature furnace (Thermal Technology 1000-2560-FP20) equipped with a water-cooling system. The diamond samples, contained in double graphite crucibles, were placed in an annealing chamber, and purged with Ar. The chamber pressure was kept at $\sim$1 mbar at the whole annealing process. The samples were then heated from room temperature to 1700\textsuperscript{o}C at a rate of 45\textsuperscript{o}C/min and held for 30 mins, followed by cooling to room temperature at a rate of 30\textsuperscript{o}C/min. Boiling acid treatment (perchloric acid : nitric acid : sulfuric acid = 1 : 1 : 1) was performed to clean the surface of the sample before any optical measurements. Optical measurements were performed by our home-built confocal microscope. The relaxation and decoherence time ($T_{1}$, $T_{2}$ and $T_{2}^{*}$) and the concentration of nearby defects (P1 and H3) were studied. Some commercial samples (raw) were also measured for a direct comparison with the “enhanced” samples (annealed). See Supplementary Note I and II for experimental details.

\bibliography{references}

\end{document}